# Improvements to enhance robustness of third-order scale-independent WENO-Z schemes

Qin Li[1a], Xiao Huang[1], Pan Yan[1], Guozhuo Tan[1], Yi Duan[2], Yancheng You[1]

[1]School of Aerospace Engineering, Xiamen University, Xiamen, Fujian, 361102, China

[2]Science and Technology on Space Physics Laboratory, China Academy of Launch Vehicle Technology, Beijing, 100076, China

**Abstract** Although there are many improvements to WENO3-Z that target the achievement of optimal order in the occurrence of the first-order critical point ($CP_1$), they mainly address resolution performance, while the robustness of schemes is of less concern and lacks understanding accordingly. In light of our analysis considering the occurrence of critical points within grid intervals, we theoretically prove that it is impossible for a scale-independent scheme that has the stencil of WENO3-Z to fulfill the above order achievement, and current scale-dependent improvements barely fulfill the job when $CP_1$ occurs at the middle of the grid cell. In order to achieve scale-independent improvements, we devise new smoothness indicators that increase the error order from 2 to 4 when $CP_1$ occurs and perform more stably. Meanwhile, we construct a new global smoothness indicator that increases the error order from 4 to 5 similarly, through which new nonlinear weights with regard to WENO3-Z are derived and new scale-independents improvements, namely WENO-$Z_{ES2}$ and -$Z_{ES3}$, are acquired. Through 1D scalar and Euler tests, as well as 2D computations, in comparison with typical scale-dependent improvement, the following performances of the proposed schemes are demonstrated: The schemes can achieve third-order accuracy at $CP_1$ no matter its location in the stencil, indicate high resolution in resolving flow subtleties, and manifest strong robustness in hypersonic simulations (e.g., the accomplishment of computations on hypersonic half-cylinder flow with Mach numbers reaching 16 and 19, respectively, as well as essentially non-oscillatory solutions of inviscid sharp double cone flow at $M = 9.59$), which contrasts the comparative WENO3-Z improvement.

**Keywords**: WENO-Z scheme; critical point; mapping method; smoothness indicator; robustness

## 1 Introduction

It is well known that WENO implementation in [1], abbreviated as WENO-JS, has been widely applied in computational fluid dynamics. The third-order version, WENO3-JS, is particularly interesting for engineers for its efficiency and robustness. Therefore, the specific improvements of WENO-JS from the view of WENO-Z [2-4] are considered in this study.

The original third-order WENO-Z, or WENO3-Z, was derived [4] by following the implementation of the fifth-order WENO-Z [2]. Reference [5] indicated that WENO3-Z would not satisfy the accuracy relation to achieve the third-order accuracy, especially in the occurrence of critical points. Several improvements were proposed to recover the optimal order in the case of first-order critical points ($f' = 0, f''\&f''' \neq 0$) under the framework of WENO3-Z, such as WENO-NP3 [6], -F3 [7], -NN3 [8], and -PZ3 [9]. As indicated in [10], although the improvements were proposed from different perspectives, the global smoothness indicators ($\tau$) therein actually assumed essentially the same form, namely, $c(f_{i-1} - 2f_i + f_{i+1})^2$, where $f$ denotes the variable and $c > 0$ is the coefficient. Moreover, Reference [11] pointed out the following: (1) WENO-NP3, -F3, and -NN3 cannot fulfill the order recovery in $L_\infty$-norm as long as the critical point occurs at the half nodes, while WENO-PZ3 even fails to achieve the order once the critical point appears; (2) In order to achieve the order recovery, the constructions of the above schemes rely on the assumption that the first-order critical point occurs at $x_j$ regarding the discretization of $(\partial f/\partial x)_j$. However, this assumption is not comprehensive and would yield the incorrect formulation in the case of WENO-PZ3; (3) All the above-mentioned improvement schemes are scale-dependent, meaning that inconsistent results will be yielded when a different variable scale or length scale is employed.

---
[a] Corresponding author. Email: qin-li@vip.tom.com

As we once indicated [12], the critical point could occur at any position within the grid cell, and consequent accuracy relations of smoothness indicators might be varied. Reference [11] further elaborated the indication and systemized the analysis considering the occurrence of critical points within grid intervals. According to the analysis in [11], we derived the correct solution for WENO-PZ3, which was validated by computation. Based on the theoretical outcomes, two scale-independent third-order schemes were devised. Both schemes expand the grid stencil(s) of WENO3-Z, and the one that only expands the downwind stencil makes use of the mapping in [12] and is referred to as WENO3-ZM. In [11], WENO3-ZM seemed more appealing because it fulfilled all tests there. Both schemes succeeded in optimal $L_\infty$-order recovery at the first-order critical point; however, in our subsequent tests where the inflow had a large Mach number, we found WENO3-ZM indicates unsatisfactory robustness (e.g., oscillations or even blow-up occurs in supersonic cylinder flows when the Mach number is larger than 4). Evidently, such issues should be analyzed and solutions be found to improve robustness, which is critical to practical applications.

In Section 2 of this paper, we first review former improvements to the third-order WENO-Z, especially the scale-independent one in [11]. Then, in Section 3, we provide new improvements with substantial robustness enhancement. Next, numerical tests are carried out in Section 4, where the robustness of a typical improvement to WENO3-Z is also comparatively addressed. Conclusions are drawn in Section 5.

## 2 Improvements to third-order WENO-Z and measures with scale-independence property

In order to facilitate discussion, the formulas of WENO are described first. Consider the one-dimensional scalar hyperbolic conservation law:

$$u_t + f(u)_x = 0, \tag{1}$$

where $\partial f(u)/\partial u > 0$. Taking the semi-discretization of Eq. (1) at $x_j$, the conservative scheme $\hat{f}(x)$ works in the form of

$$(f(u)_x)_j \approx (\hat{f}_{j+1/2} - \hat{f}_{j-1/2})/\Delta x. \tag{2}$$

The formulation of WENO-JS [1] is

$$\hat{f}_{j+1/2} = \sum_{k=0}^{r-1} \omega_k q_k^r \text{ with } q_k^r = \sum_{l=0}^{r-1} a_{kl}^r f(u_{j-r+k+l+1}), \tag{3}$$

where $r$ is the grid number of candidate scheme $q_k^r$ with coefficient $a_{kl}^r$ included, and $\omega_k$ is the normalized nonlinear weight corresponding to the linear counterpart $d_k$. $\omega_k$ is usually evaluated from the non-normalized weight $\alpha_k$ by $\omega_k = \alpha_k / \sum_{l=0}^{r-1} \alpha_l$. To derive $\alpha_k$, a smoothness indicator should be applied, and the canonical ones by Jiang and Shu [1] can be defined in positive semi-definite quadratic form as:

$$\beta_k^{(r)} = \sum_{m=0}^{r-2} c_m^r \left( \sum_{l=0}^{r-1} b_{kml}^r f(u_{j-r+k+l+1}) \right)^2. \tag{4}$$

The coefficients $a_{kl}^r$, $b_{kml}^r$, and $c_m^r$ can be found in [11-12]. For example, $\beta_0^{(2)} = (f_j - f_{j-1})^2$ and $\beta_1^{(2)} = (f_{j+1} - f_j)^2$. Making use of $\beta_k^{(r)}$, $\alpha_k$ in WENO-JS is defined as: $\alpha_k = d_k/\left(\varepsilon + \beta_k^{(r)}\right)^2$ [1] where $\varepsilon = 10^{-6} \sim 10^{-7}$; while for WENO-Z, $\alpha_k$ may take the form [2-4]

$$\alpha_k = d_k \left( 1 + c_\alpha \left( \frac{\tau}{\beta_k^{(r)} + \varepsilon} \right)^p \right), \tag{5}$$

where $\tau$ is the global smoothness indicator, $p = 1$ or 2, $c_\alpha$ may take 1, and $\varepsilon$ can take a small value such as $10^{-40}$. As indicated in [11] (see Proposition 3 there), a smaller $c_\alpha$ would benefit resolution but conversely be unfavorable for stability.

From [2, 13], in order to achieve the optimal $(2r-1)$th-order, $\omega_k$ should satisfy the necessary and sufficient conditions:

$$\begin{cases} \sum_{k=0}^{r-1} A_k(\omega_k^+ - \omega_k^-) = O(\Delta x^r) \\ \omega_k^\pm - d_k = O(\Delta x^{r-1}) \end{cases}, \tag{6}$$

where the superscript "$\pm$" corresponds to the location $x_{j\pm\frac{1}{2}}$, and $A_k$ is the relevant coefficient, or satisfy the sufficient condition:

$$\omega_k^\pm - d_k = O(\Delta x^r). \tag{7}$$

In the absence of critical points, WENO-JS is verified to have Eq. (6) established [2]. In the occurrence of a critical point which is usually assumed to locate at $x_j$ [2-3, 6-9, 13], order degradation occurs. In case of WENO3-Z, where $\tau = \tau_3 = \left|\beta_1^{(2)} - \beta_0^{(2)}\right| = |(f_{j+1} - f_{j-1})(f_{j+1} - 2f_j + f_{j-1})|$ and $p = 2$ in Eq. (5), the improvements to recover optimal order mainly fall in two classes, namely, attempts to satisfy Eq. (7) [6-7] or Eq. (6) [8-9]. The corresponding $\alpha_k$ usually takes the form of

$$\alpha_k = d_k \left( 1 + \frac{\tau^{p_1}}{\left(\beta_k^{(r)} + \varepsilon\right)^{p_2}} \right), \tag{8}$$

where $p_1 \neq p_2$, and $\tau$ can be expressed as

$$\tau = \begin{cases} c_{\tau_1} |(f_{j+1} - f_{j-1})(f_{j+1} - 2f_j + f_{j-1})| & (9.1) \\ c_{\tau_2} (f_{j+1} - 2f_j + f_{j-1})^2 & (9.2) \end{cases}.$$

By means of Eqns. (8) and (9), the aforementioned improvements to WENO3-Z can be reproduced [11] by specific choices of $\{p_1, p_2, c_{\tau_1}, c_{\tau_2}\}$ as shown in Table 1.

Table 1 Parameters in Eqns. (8) and (9) of improvements to WENO3-Z

| Schemes | $\alpha_k$ by Eq. (9) | | $\tau$ | |
|---|---|---|---|---|
| | $p_1$ | $p_2$ | Eq. (9.1) with $c_{\tau_1}$ as | Eq. (9.2) with $c_{\tau_2}$ as |
| WENO-NP3 [6] | 3/2 | 1 | -- | 10/12 |
| WENO-F3 [7] | 3/2 | 1 | -- | 2/12 |
| WENO-NN3 [8] | 1 | $\leq 3/4$ | -- | 10/12 |
| WENO-PZ3 [9] | 1 | $\leq 1/2$ | 1 | -- |

As indicated in [11], improvements in Table 1 are scale-dependent, or their $\alpha_k$ contain a dimension regarding $[f]$, and the optimal order is NOT achieved when the critical points occur on half nodes; moreover, the assumption that critical points occur at $x_j$ is inappropriate in some cases, with ensuing incorrect solutions, such as the incorrect value of $p_2$ for WENO-NN3 (the correct solution should be $p_2 \leq 1/2$) [12]. Based on the understandings first proposed in [12], we elaborated the analysis considering the occurrence of critical points within grid intervals, derived corresponding smoothness indicators, and proposed a scale-independent WENO-Z scheme with incorporation of mapping in [11]. Because such works are closely related with the investigation in Section 3, a brief review is given below.

(1) Analysis considering the occurrence of critical points within grid intervals

Considering WENO3-Z, where $r = 2$ in Eq. (2), one can see the whole stencil is $\{x_{j-1}, x_j, x_{j+1}\}$. When a critical point occurs, its location is supposed to span the stencil with the coordinate as:

$$x_c = x_j + \lambda \cdot \Delta x \text{ where } -1 < \lambda < 1. \tag{10}$$

Based on Eq. (10), the Taylor expansion can be expressed in terms such as $\beta_k^{(r)}$ and $\tau$ regarding $x_c$. It can be seen that the coefficients of leading error of $\beta_k^{(2)}$ and some $\tau$ vary with $\lambda$ when first-order critical points occur (e.g., the leading error of $\beta_{0,1}^{(2)}$ is $\frac{1}{4}(2\lambda \pm 1)^2 f_{x_c}''^2 \Delta x^4$, and that of $\tau_3$ is $|2\lambda f_{x_c}''^2 \Delta x^4|$). Therefore, when $\lambda = \pm\frac{1}{2}$, the accuracy order of one $\beta_k^{(2)}$ will become 6, and when $\lambda = 0$, that of $\tau_3$ will become 5. More information of other indicators is provided in [11-12].

Because $\beta_k^{(2)}$ has the dimension of $[f]^2$, $\tau$ should have the same dimension to make $\tau/\beta_k^{(2)}$ dimensionless and be quadratic thereby. In [11], a proposition, namely Proposition 5 therein, was proposed for $\tau$, which is defined on $\{x_{j-1}, x_j, x_{j+1}\}$ as follows:

**Proposition** [11]. Consider the generic quadratic form of $f$ as $\tau(f) = (f_{j-1}, f_j, f_{j+1})[a_{i_1 i_2}](f_{j-1}, f_j, f_{j+1})^T$, where $[a_{i_1 i_2}]$ is a $3 \times 3$ matrix and $i_1, i_2 = 1, \ldots, 3$. (a) Assuming that the first-order critical point occurs at $x_c = x_j + \lambda \cdot \Delta x$, where $-1 < \lambda < 1$ and $\{f'_{x_c} = 0, f''_{x_c} \& f'''_{x_c} \neq 0\}$, a nontrivial solution of $a_{i_1 i_2}$ does not exist such that the Taylor expansion of $\tau(f)$ toward $x_c$ has the leading error $O(\Delta x^5)$. (b) When noncritical points occur, the only form that $\tau(f)$ can take is $c(f_{j+1} - 2f_j + f_{j-1})^2$ if it has a leading error of $O(\Delta x^4)$.

In the following, the first-order critical point (i.e., $f' = 0, f'' \& f''' \neq 0$) is abbreviated as $CP_1$ for brevity. It can be proven that the statement in part "(b)" holds in the case of first-order critical points as well. Hence, one can see any improvements who claim to have $\tau$ with an error of $O(\Delta x^4)$ (e.g., WENO-NP3, -F3, and -NN3) are actually of the same kind in using Eq. (9.2) but with different $c_{\tau_2}$; moreover, it is impossible to derive a $\tau$ with an error of $O(\Delta x^{\geq 5})$ for the purpose of satisfying Eq. (7) under the use of $\beta_k^{(2)}$ unless the stencil be expanded, or it is unavailable for any scale-independent scheme to achieve the third-order at $CP_1$ by $\beta_k^{(2)}$ and $\tau$ on $\{x_{j-1}, x_j, x_{j+1}\}$. In view of the above, an expansion in [11] was chosen as $\{x_{j-1}, x_j, x_{j+1}, x_{j+2}\}$, which will be introduced subsequently.

(2) Construction of $\tau$ satisfying accuracy requirement

As indicated in [11], when $\lambda = \pm(1/2)$, the leading terms of $\beta_{0,1}^{(2)}$ become $O(\Delta x^6)$, and it is hardly likely to derive a $\tau$ on the expended stencil with an error of $O(\Delta x^{\geq 7})$ in two situations. In view of this, $\beta_1^{(2)}$ was extended [11] to $\beta_2^{(3)}$ also such that $\tau$ is not required to be of $O(\Delta x^{\geq 7})$ at $\lambda = 1/2$. Correspondingly, a unique indicator $\tau_{cp1}$ was derived as [11]:

$$\tau_{cp1} = c \times \left|(-f_{j+2} + 3f_{j+1} + 21f_j - 23f_{j-1}) \times (f_{j+2} - 3f_{j+1} + 3f_j - f_{j-1})\right|, \tag{11}$$

where its Taylor expansion at $f'_{xc} = 0$ is $|(-6\lambda - 3)f''_{xc}f'''_{xc}\Delta x^5 + O(\Delta x^6)|$ when $\lambda \neq -1/2$ and otherwise $O(\Delta x^7)$. By means of $\tau_{cp1}$, $\tau_{CP_1}/\beta_0^{(2)}$ and $\tau_{CP_1}/\beta_2^{(3)}$ would have $O(\Delta x^n)$ with $n \geq 1$ in the case of $CP_1$, and thereby Eq. (7) would be satisfied after using Eq. (5) with $p = 2$.

(3) Incorporation of mapping

Reference [2] indicated that the constant use of $p = 2$ was liable to less resolution in problems such as Shu–Osher cases. In view of this, a newly developed rational mapping method in [12] was incorporated with Eq. (5) [11] such that

$$\alpha_k = d_k(1 + M(\tau/\beta_k)), \tag{12}$$

where $M(\cdot)$ is the mapping function. For the third-order scheme, $M(\cdot)$ is specialized as

$$M(\omega) = \begin{cases} \frac{\omega^2}{\omega + c_2\omega(c_3-\omega)^2 + c_1(c_3-\omega)^2}, & \omega \leq c_3 \\ \omega, & \omega > c_3 \end{cases}, \tag{13}$$

where $\{c_1, c_2, c_3\}$ takes $\{1.2, 0.1, 55\}$ for $d_0 = 1/3$ and $\{1.2, 0.1, 35\}$ for $d_1 = 2/3$. It is shown in [11] that $M(\omega)$ has the following properties:

$$M(0) = M'(0) = 0, \ M''(0) \neq 0; \ M(c_3) = c_3, \ M'(c_3) = 1. \tag{14}$$

Thus far, the final scheme WENO3-ZM is accomplished by Eqns. (12), (13), and (11) and $\beta_k = \{\beta_0^{(2)}, \beta_2^{(3)}\}$. Apparently, WENO3-ZM is scale-independent because the argument $\tau/\beta_k$ in $M(\cdot)$ is dimensionless. The parameters $\{c_1, c_2, c_3\}$ are so chosen that WENO3-ZM has resolutions outperforming WENO-NP3, -F3, -NN3, and -PZ3, and fulfills the tests in [11] (e.g., 1D strong shock wave, blast wave, Shu–Osher problems, 2D Riemann problems, and double Mach reflection).

Although WENO3-ZM can achieve optimal order recovery in cases of $CP_l$, more follow-up tests reveal that it is less robust (e.g., in case of a supersonic cylinder, the computations become oscillatory or blow up when $M > 4$). Similarly, we find the same insufficiency exists in the improvements in Table 1. In view of this, we further analyze and propose new implementations that can work at $M \geq 16$ while preserving the properties such as scale-independence, optimal order recovery at $CP_l$, and high resolutions.

**3 Improvements to enhance robustness based on stencil expansion**

First, a heuristic analysis is provided for the sake of enhancing the robustness of WENO-Z improvements such as WENO3-ZM while achieving optimal order at critical points.

(1) Heuristic analysis on the robustness of WENO3-ZM

As indicated in Section 3, when $CP_l$ occurs at (or approaching) $\lambda = -1/2$, the error of $\tau_{cp1}$ would become $\left|-\frac{1}{4}f'''_j f_j^{(4)}\Delta x^7 + O(\Delta x^8)\right|$ from the original $O(\Delta x^5)$. This change might make $\tau_{cp1}/\beta_k$ have a suddenly small magnitude and subsequently make $M(\tau/\beta_k)$ in Eq. (13) smaller than 1 overall. If such a situation occurs near discontinuities such as a strong shock wave, the evaluation might affect the ENO (essentially non-oscillation) property, and thereby oscillations will arise and lead to blow-up eventually. Empirical support for this viewpoint is that if Eq. (13) is modified as $\alpha_k = d_k(1 + c_\alpha \times M(\tau/\beta_k))$, where $c_\alpha$ would take a considerably large positive number, WENO3-ZM can fulfill the computation of supersonic cylinder flows at large Mach numbers (e.g., the computation at $M = 15$ can be accomplished providing $c_\alpha = 10^{15}$). However, such a choice will make the scheme rather dissipative, perform poorly in terms of rate of numerical convergence at critical points, and indicate inferior resolutions in Shu–Osher problems.

As shown in Section 2, the motivation to develop $\tau_{cp1}$ in WENO3-ZM originated from the difficulty caused by very small $\beta_0^{(2)}$ near $CP_I$, namely, $O(\Delta x^6)$ at $\lambda = -1/2$. In view of the expansion of $\beta_1^{(2)}$ to $\beta_2^{(3)}$ there, the same practice is considered here for $\beta_0^{(2)}$, through which $\beta_0^{(2)}$ can also have the magnitude $O(\Delta x^4)$ at $CP_I$, and tiny errors such as $O(\Delta x^6)$ can be avoided. If both $\beta_k$ values have a magnitude of $O(\Delta x^4)$ at $CP_I$, we only need a $\tau$ with at most $O(\Delta x^5)$ to accommodate Eq. (5) with $p = 2$, through which the sufficient condition Eq. (7) is satisfied. As just implied, $\tau$ with less order is assumed to favor stability and robustness; hence, such a choice is worthy of analysis and practice.

In short, the following aspects need to be investigated on the expanded stencil: (a) The formulation of $\tau$ that satisfies the requirement of accuracy relation while favoring robustness as much as possible; (b) The optimal formulation of $\beta_k$, which is of $O(\Delta x^2)$ in the absence of critical points and of $O(\Delta x^4)$ at $CP_I$. Corresponding solutions are given below.

(2) $\tau$ with the error $O(\Delta x^5)$ at $CP_I$

According to [1], the following understanding is widely accepted: $\{\beta_k\}$, which have the dimension of $[f]^2$, would contain the contribution of first-order derivatives and therefore have the magnitude of $O(\Delta x^2)$ in the absence of critical points, and then $\tau$ should at least be of $O(\Delta x^{\geq 3})$ accordingly. Reference [12] indicated that upon construction of a scale-independent scheme, $\tau$ is essentially comprised of the multiplication (or corresponding combinations) of two undivided discretizations of derivatives regarding $x_j$. Hence, in order to obtain a multiplication with error as $O(\Delta x^5)$ when $CP_I$ occurs, at least one discretization of the third-order derivative should be included. Based on the understanding in "(1)," a global indicator is chosen as: $\tau = |\delta_j^{(1)} \delta_j^{(3)}|$, where $\delta_j^{(n)}$ denotes the undivided approximation of the $n$th-order derivative with regard to $x_j$. One can see the $\delta_j^{(3)}$ on stencil $\{x_{j-1}, x_j, x_{j+1}, x_{j+2}\}$ is specified as

$$\delta_j^{(3)_1} = (f_{j+2} - 3f_{j+1} + 3f_j - f_{j-1}), \tag{15}$$

where $m$ in $\delta_j^{(n)m}$ denotes the accurate order of $\delta_j^{(n)}/\Delta x^n$ on approximating $(\partial f/\partial x)_j$. Considering that the explicit dependence of the third-order scheme in Eq. (3) is $\{x_{j-1}, x_j, x_{j+1}\}$, the same stencil is chosen for $\delta_j^{(1)_1}$, and its form is derived as $a_1 f_{j-1} + (-2a_1 - 1)f_j + (a_1 + 1)f_{j+1}$ with $a_1$ as the free parameter. Supposing $CP_I$ occurs within the stencil, the scope of $x_c$ or the range of $\lambda$ should be $\lambda \in [-1,1]$ or $\left(-\lambda + \frac{1}{2}\right) \in \left[-\frac{1}{2}, \frac{3}{2}\right]$, and the accuracy relation of $\delta_j^{(1)_1}$ can be derived as $\left(a_1 - \lambda + \frac{1}{2}\right) f''_{x_c} \Delta x^2 + O(\Delta x^3)$. Hence, the leading second-order term exists, providing $a_1 > 1/2$. In this study, the simple $a_1 = 1$ is employed, and $\delta_j^{(1)_1} = 2f_{j+1} - 3f_j + f_{j-1}$. Thus far, a new global indicator $\tau_4$ is defined as

$$\tau_4 = |(2f_{j+1} - 3f_j + f_{j-1})(f_{j+2} - 3f_{j+1} + 3f_j - f_{j-1})|. \tag{16}$$

The accuracy relations of $\tau_4$ are $|f'_j f_j^{(3)} \Delta x^4 + O(\Delta x^5)|$ when no critical points occur and $(\frac{3}{2} - \lambda) f''_{x_c} f_{x_c}^{(3)} \Delta x^5 + O(\Delta x^6)$ at $CP_I$. One may wonder about the employment of $\delta_j^{(1)}$ having a higher order and occupying the extended stencil $\{x_{j-1}, x_j, x_{j+1}, x_{j+2}\}$. In the view of this, $\delta_j^{(1)_3}$ can be chosen as $\delta_j^{(1)_3} = \frac{1}{6}(-f_{j+2} + 6f_{j+1} - 3f_j - 2f_{j-1})$, and the accuracy relation of which at $CP_I$ is: $-\lambda f''_{x_c} \Delta x^2 + \frac{1}{2} \lambda^2 f_{x_c}^{(3)} \Delta x^3 - \frac{1}{12}(2\lambda^3 + 1) f_{x_c}^{(4)} \Delta x^4 + O(\Delta x^5)$. Hence, when $\lambda = 0$, the error order of $\delta_j^{(1)_3}$ will increase from 2 to 4, which yields a much smaller $(\tau/\beta_k)$, and such a situation would be unfavorable for stability according to the previous discussion. In the next "(4)," numerical supports are provided to justify the statement by comparing $\tau_4$ with other candidates on robustness.

(3) Formulations of $\beta_k^{(2)}$ on expanded stencil

As shown in Section 2, the extension of $\beta_1^{(2)}$ to $\beta_2^{(3)}$ lowers the order of error at $CP_I$. Seeing that $\beta_2^{(3)} = \frac{1}{4}(3f_i - 4f_{i+1} + f_{i+2})^2 + \frac{13}{12}(f_i - 2f_{i+1} + f_{i+2})^2$, the first part on the left regards the discretization of $(\frac{\partial f}{\partial x})_j^2$ as $\beta_1^{(2)}$ but with one order higher; however, such discretization increases grid dependence, which might be unfavorable to robustness. To justify the supposition, numerical supports are also provided in the next "(4)." For considering this, a different expansion of $\beta_1^{(2)}$ is employed as follows:

$$\beta_1^{(2)*} = \beta_1^{(2)} + c_{\beta_1} \left(\delta_{j+1}^{(2)_2}\right)^2, \tag{17}$$

where $c_{\beta_1}$ is a positive parameter, and $\delta_{j+1}^{(2)_2} = f_{j+2} - 2f_{j+1} + f_j$. Our numerical experiments indicate that a larger $c_{\beta_1}$ is apt to weaken numerical stability but increase the convergence rate of numerical order, and the recommended value is $c_{\beta_1} = 0.15$.

Likewise, it is wondered whether similar procedures be casted toward $\beta_0^{(2)}$ to mitigate the same difficulty. Apparently, the candidates for the stencil expansion of an indicator would naturally be $\{x_{j-1}, x_j, x_{j+1}\}$ or $\{x_{j-2}, x_{j-1}, x_j\}$.

(3.1) Extension of $\beta_0^{(2)}$ on the stencil $\{x_{j-1}, x_j, x_{j+1}\}$

In view of minimizing the stencil dependence, the extension of $\beta_0^{(2)}$ still employs the form

$$\beta_0^{(2)*} = \beta_0^{(2)} + c_{\beta_0} \left(\delta_j^{(2)_2}\right)^2, \tag{18}$$

where $\delta_j^{(2)_2} = (f_{j+1} - 2f_j + f_{j-1})$ and $c_{\beta_0} > 0$. Equation (18) indicates the whole stencil of $\beta_k^*$ remains unchanged as $\{x_{j-1}, x_j, x_{j+1}, x_{j+2}\}$. The accuracy relation of $\beta_k^{(2)*}$ at $CP_I$ can be derived as $\left((\lambda + \frac{1}{2})^2 + c_{\beta_0}\right) f_{x_c}''^2 \Delta x^4 + O(\Delta x^5)$. One can see that Eq. (18) would recover the optimal order when Eq. (16) and Eq. (5) with $p = 2$ are used. However, because the stencil expansion overlaps with that of $\beta_1^{(2)*}$, the computation contains shock waves and will blow up unless $c_{\beta_0} < 10^{-4} \sim 10^{-6}$. Our numerical tests indicate, under such choice of $c_{\beta_0}$, the corresponding scheme would barely achieve third-order in Case 1 in Section 4.1 where $CP_I$ would locate on the half node occasionally; meanwhile, if the optimal order be achieved in such a case, the value of $c_{\beta_0}$ should satisfy $c_{\beta_0} > 10^{-1} \sim 10^{-2}$. Hence, a distinct gap of $c_{\beta_0}$ exists in both cases, which indicates it should be evaluated nonlinearly (i.e., being nearly zero in case of discontinuities otherwise having an appropriately small value). Apparently, an accurate detector to discern discontinuity is requisite.

As is well known, a discontinuity or shock detector is another topic of concern for investigations, such as those of Harten [15] and Jameson [16]. The normalized "weight" without $d_k$ in WENO, $(\frac{\alpha_k}{d_k})/\sum_i(\frac{\alpha_i}{d_i})$, actually bears a kind of indicator. The main focus of detectors is how to avoid misjudging the resolvable oscillation from the spurious one and discontinuity. In [14], a method is proposed from the perspective of the reduced wave number $\kappa = k\Delta x$ of variable fluctuation, where $k$ is the wave number. As an illustration [14], considering a distribution of $f(x)$ as $f = sin(kx)$, one can see $k = \sqrt{f^{(p+2)}/f^{(p)}}$ and the corresponding undivided discretization would yield $\kappa \approx \sqrt{\delta_{x^*}^{(p+2)}/\delta_{x^*}^{(p)}}$ where $x^*$ denotes some reference point. Considering $CP_I$ would make $\delta^{(1)}$ be near zero, the following formula was suggested [14] to evaluate the numerical $\kappa'$:

$$\kappa' = \sqrt{\left(\left|\delta_{x^*}^{(3)}\right| + \left|\delta_{x^*}^{(4)}\right|\right)/\left(\left|\delta_{x^*}^{(1)}\right| + \left|\delta_{x^*}^{(2)}\right| + \varepsilon\right)} \tag{19}$$

where $\varepsilon = 10^{-3}$. Reference [14] further indicated that the appropriate threshold $\kappa_c$ enables the identification of resolvable fluctuations from spurious ones and discontinuity by checking whether $\kappa' < \kappa_c$. In the context of Eq. (2), it is suggested [14] $x^*$ would take $x_{j+1/2}$ and $\kappa_c \approx 1$. In addition, the inclusion of $\delta_{x^*}^{(4)}$ indicates the stencil has at least five points and is beyond the current $\{x_{j-1}, \dots, x_{j+2}\}$.

Applying the detector via Eq. (19), a dynamic $c_{\beta_0}$ can be defined as follows, which shifts between two thresholds $c_{\beta_0}^{(0)}$ and $c_{\beta_0}^{(1)}$ with $c_{\beta_0}^{(0)} \ll c_{\beta_0}^{(1)}$, where $c_{\beta_0}^{(0)}$ suits the case of discontinuity and $c_{\beta_0}^{(1)}$ corresponds to smooth variable distribution; in addition, the thresholds are suggested by numerical experiments to be $c_{\beta_0}^{(0)} = 10^{-8}$ and $c_{\beta_0}^{(1)} = 1$.

(a) Compute the numerical $\kappa'$ by Eq. (19). Instead of evaluating $\delta^{(1)} \sim \delta^{(4)}$ at $x_{j+1/2}$ [14], we define them at $x_j$ for the sake of upwind preference. As just mentioned, the overall stencil further extends to $\{x_{j-2}, \dots, x_{j+2}\}$ because of $\delta_j^{(4)}$, and the corresponding undivided discretizations are $\delta_j^{(1)} = \frac{1}{12}(f_{j-2} - 8f_{j-1} + 8f_{j+1} - f_{j+2})$, $\delta_j^{(2)} = \frac{1}{12}(f_{j-2} - 16f_{j-1} + 30f_j - 16f_{j+1} + f_{j+2})$, $\delta_j^{(3)} = \frac{1}{2}(f_{j-2} - 2f_{j-1} + 2f_{j+1} - f_{j+2})$, and $\delta_j^{(4)} = (f_{j-2} - 4f_{j-1} + 6f_j - 4f_{j+1} + f_{j+2})$. In Euler equations, the variable used in the scheme usually employs the characteristic form, and the acquisition of which at $x_{j-2}$ implies extra computations. In view of this, future optimization will be practiced to reduce computation costs, such as choosing a scalar variable to derive $\kappa'$.

(b) Compute the reference indicator $\psi \in [0,1]$ based on indicators of WENO3-Z. Given the well-established performance of WENO3-Z despite its order degradation at critical points, a preliminary indicator is derived first as

$$\psi_z = 1 - \frac{\tau_3}{\beta_0^{(2)} + \beta_1^{(2)} + \epsilon_\psi}, \tag{20}$$

where $\tau_3 = \left|\beta_1^{(2)} - \beta_0^{(2)}\right|$ [4] and $\epsilon_\psi = 10^{-40}$. $\psi_z$ is expected to approach 0 at discontinuity and 1 at the smooth region although it may misjudge the local extrema for the discontinuity, as described previously. Using $\psi_z$ and the amplifier factor $1/\psi_c > 1$, $\psi$ is defined as $\psi = \min(1, \frac{\psi_z}{\psi_c})$. The larger the $\psi_c$, the more robustness benefits; comprehensively, $\psi_c = 0.3$.

(c) Construct the adaptor $\sigma$ to evaluate discontinuity by using $\kappa'$ and $\psi$. As shown in [14], a switch of detection can be constructed as: $\frac{1+sign(k_c - k)}{2} + \frac{1-sign(k_c - k)}{2}$ with the use of threshold $\kappa_c$. Accordingly, $\sigma$ can be defined as:

$$\sigma = \frac{1+sign(k_c - k)}{2} + \frac{1-sign(k_c - k)}{2}\psi, \tag{21}$$

where the discontinuity strength is represented by $\psi$. It is worth mentioning that in order to have a scheme accomplishing the computation with Mach numbers as high as 20, $\kappa_c$ would be smaller than that suggested by [14] (i.e., 0.75).

(d) Compute $c_{\beta_0}$ adaptively. $c_{\beta_0}$ is defined nonlinearly as

$$c_{\beta_0} = \left(c_{\beta_0}^{(0)} + \sigma^{p_\sigma}\left(c_{\beta_0}^{(1)} - c_{\beta_0}^{(0)}\right)\right), \tag{22}$$

where $p_\sigma$ acts to enhance the stability of the scheme and takes 2 temporarily.

Thus far, $\beta_k^{(2)*}$ have been defined, whose accuracy is $O(\Delta x^2)$ at non-critical points and becomes $O(\Delta x^4)$ at $CP_l$, and therefore that of $\tau_4/\beta_k^{(2)*}$ would be $O(\Delta x^2)$ and $O(\Delta x)$. Considering the notation of improvements to WENO3-Z based on expanding the stencil as WENO3-$Z_{ES}$ [11], the above improvement is referred to as WENO3-$Z_{ES2}$. To facilitate coding, its implementation is summarized as:

(a) Compute $\beta_k^{(2)}$ by Eq. (4) and $\tau_3 = \left|\beta_1^{(2)} - \beta_0^{(2)}\right|$.

(b) Compute $\psi_z$ by Eq. (20), then compute $\psi = \min(1, \frac{\psi_z}{\psi_c})$ where $\psi_c = 0.3$.

(c) Compute the adaptor $\sigma$ by Eq. (21) where $\kappa_c = 0.75$, then compute $c_{\beta_0}$ by Eq. (22)

where $c_{\beta_0}^{(0)} = 10^{-8}$, $c_{\beta_0}^{(1)} = 1$, and $p_\sigma = 2$.

(d) Using Eqns. (17)-(18) together with the obtained $c_{\beta_0}$ and $c_{\beta_1} = 0.15$, the extended $\beta_k^{(2)*}$ are obtained.

(e) Compute $\tau_4$ by Eq. (16), then use Eq. (5) with $c_\alpha = 0.15$ and $p = 2$ to derive $\alpha_k$.

(f) Compute $\omega_k$ by normalizing $\alpha_k$, and $\hat{f}_{j+1/2}$ is acquired by using Eq. (3) at last.

(3.2) Extension of $\beta_0^{(2)}$ on the stencil $\{x_{j-2}, x_{j-1}, x_j\}$

Still in light of minimizing stencil dependence as much as possible, the extension takes the following form:

$$\beta_0^{(2)*} = \beta_0^{(2)} + c_{\beta_0}\left(\delta_{j-1}^{(2)_2}\right)^2, \tag{23}$$

where $\delta_{j-1}^{(2)_2} = \left(f_j - 2f_{j-1} + f_{j-2}\right)^2$. One can see that the overall stencil of Eq. (23) does not intersect with that of Eq. (15), which indicates $\beta_0^{(2)*}$ and $\beta_1^{(2)*}$ could discern the discontinuity distinctly, and therefore the linear definition of $c_{\beta_0}$ would make the scheme achieve the ENO property and be free of instability. However, our numerical experiments indicate that a too-large $c_{\beta_0}$ would yield numerical instability, whereas a too-small $c_{\beta_0}$ would impair the rate of order convergence. The recommended value of $c_{\beta_0}$ is 0.6. For convenience, the improvement is referred to as WENO3-$Z_{ES3}$, and its implementation is summarized as:

(a) Compute the extended $\beta_k^{(2)*}$ by Eqns. (17) and (23) where $c_{\beta_0} = 0.6$ and $c_{\beta_1} = 0.15$.

(b) Compute $\tau_4$ by Eq. (16), then use Eq. (5) with $c_\alpha = 0.4$ and $p = 2$ to derive $\alpha_k$.

(c) Compute $\omega_k$ by normalizing $\alpha_k$, and $\hat{f}_{j+1/2}$ is acquired by using Eq. (3) at last.

(4) Numerical supports to justify the definition of $\beta_1^{(2)*}$ and $\tau_4$

In the above discussion, the employment of $\beta_1^{(2)*}$ as an extension of $\beta_1^{(2)}$ other than $\beta_2^{(3)}$ and the choice of $\tau_4$ as $\tau$ arise from the perspective of robustness, whose rationality is supported by numerical tests as follows.

(a) Employment of $\beta_1^{(2)*}$ other than $\beta_2^{(3)}$

A specific WENO3-Z by Eq. (5) is chosen as the reference, where $p = 2$ and $c_\alpha = 0.02$. The following two variants of the scheme are derived with the following modifications: the first one uses $\beta_1^{(2)*}$ by Eq. (17) to replace $\beta_1^{(2)}$, where $c_{\beta_1} = 0.15$, and the second one uses $\frac{1}{4}(3f_i - 4f_{i+1} + f_{i+2})^2 + c_{\beta_1}^*(f_i - 2f_{i+1} + f_{i+2})^2$ as the substitute, where $c_{\beta_1}^* = 13/12$. The double Mach reflection is numerically tested using the conditions in Section 4.1, where the first variant fulfills the computation whereas the second one fails, and further computations where the second scheme is used with $c_{\beta_1}^* \in [0, 13/12]$ blow up also. Hence, in the extension of $\beta_1^{(2)}$, the employment of $(f_{i+1} - f)$ to discretize $\left(\frac{\partial f}{\partial x}\right)_j$ performs more robustly than that of $\frac{1}{2}(3f_i - 4f_{i+1} + f_{i+2})$.

(b) Employment of $\tau_4$ instead of other candidates

As shown in "(2)," there is a candidate $\tau$ such that $\tau/\beta_k$ is of $O(\Delta x^{\geq 1})$ in the presence of

$CP_l$ or $O(\Delta x^{\geq 2})$ in the absence of critical points providing that $\beta_k$ is of $O(\Delta x^4)$ or $O(\Delta x^2)$ respectively (e.g., $\left|\delta_j^{(1)_3}\delta_j^{(3)_1}\right|$ herein). As shown in [11], one can check similar candidate $\tau$ values exist such as $\left|\delta_j^{(2)_2}\delta_j^{(3)_1}\right|$ and $\left(\delta_j^{(3)_1}\right)^2$, where $\delta^{(2)_2} = (f_{j+1} - 2f_j + f_{j-1})$, and the accuracy of $\{\left|\delta_j^{(1)_3}\delta_j^{(3)_1}\right|, \left|\delta_j^{(2)_2}\delta_j^{(3)_1}\right|, \left(\delta_j^{(3)_1}\right)^2\}$ are of $O(\Delta x^4)$, $O(\Delta x^5)$, and $O(\Delta x^6)$ in the absence of critical points and of $O(\Delta x^5)$, $O(\Delta x^5)$, and $O(\Delta x^6)$ at $CP_l$ when $\lambda \neq 1/2$, respectively. To justify the employment of $\tau_4$ values other than the above candidates, the following tests are carried out: Still starting from $\alpha_k = d_k \cdot (1 + 0.02\left(\tau/\beta_k^{(2)}\right)^2)$; concerning the analysis in "(a)" and the upcoming application of $\beta_1^{(2)*}$, substitute $\beta_1^{(2)*}$ for $\beta_1^{(2)}$ and use the above candidates as well as $\tau_4$ as $\tau$, and thereafter the final test schemes are acquired; after that, the double Mach reflection is checked. Although all test schemes can fulfill the computation, numerical oscillations occur in the results of schemes where $\tau$ employs $\left|\delta_j^{(1)_3}\delta_j^{(3)_1}\right|$ or $\left(\delta_j^{(3)_1}\right)^2$, as shown in Fig. 1, whereas the results of schemes where $\tau$ uses $\tau_4$ or $\left|\delta_j^{(2)_2}\delta_j^{(3)_1}\right|$ are free of oscillations. Hence, the former $\tau$ values that have a higher order of error in the presence or absence of $CP_l$ numerically indicate less robustness. Furthermore, in the following case of hypersonic cylinder flow in Section 4.2, if $\tau_4$ in WENO3-$Z_{ES2}$ is replaced by $\left|\delta_j^{(2)_2}\delta_j^{(3)_1}\right|$, the computation can be accomplished only if $M \leq 14$, whereas the computation of WENO3-$Z_{ES2}$ succeeds with higher Mach numbers as 16. In short, $\tau_4$ indicates more robustness than other candidates, which coincides with the analysis in "(2)."

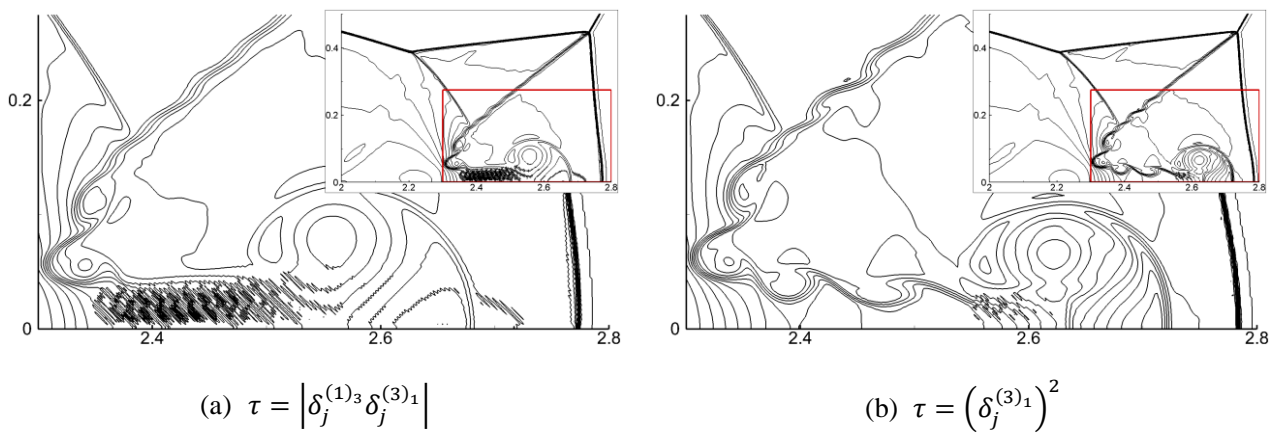

(a) $\tau = \left|\delta_j^{(1)_3}\delta_j^{(3)_1}\right|$  (b) $\tau = \left(\delta_j^{(3)_1}\right)^2$

Fig. 1. Numerical oscillations in double Mach reflection by test schemes employing different $\tau$

## 4 Numerical examples

4.1 Case descriptions

Three kinds of equations are considered: the equation of 1D scalar advection, 1D Euler equations, and 2D Euler/Navier-Stokes equations.

(1) 1D scalar advection equation

The governing equation is $\partial u/\partial t + \partial u/\partial x = 0$ with various initial conditions $u(x, 0)$ corresponding to specific problems. The following initial condition is especially chosen as:
$$u(x,0) = \sin(\pi(x - x_c) - \sin(\pi(x - x_c))/\pi), \tag{24}$$
where $x \in [-1,1]$ and $x_c = 0.5966831869112089637212$. $x_c$ is so chosen that $u(x, 0)$ would have two $CP_1$ at $x = 0$ and $x = -2 + 2x_c$. The fourth-order Runge–Kutta (RK4) scheme is used for time discretization, and $\Delta t$ is defined as $\Delta t = CFL \cdot \Delta x$ due to $\Delta t < \Delta x^{\frac{3}{4}}$. A series of grids with numbers {10, 20, 40, 80 …} are used such that $x = 0$ initially coincides with certain grid points, and the computations run until $t = 2$ with $CFL = 0.25$. One can see that the critical point initially at $x = 0$ theoretically moves to the half-node ($\lambda = 1/2$) after every four iterations. This case is designed to check the convergence rate of WENO3-Z improvements at $CP_l$, and as shown in [11], all improvements in Table 1 failed to achieve the third order in $L_\infty$-norm.

(2) 1D Euler equations

Three problems are chosen: a strong shock wave, a blast wave, and the Shu–Osher problem.

(a) Strong shock wave

The initial condition is $(\rho, u, p) = \{(1, 0, 0.1PR), -5 \le x < 0; (1, 0, 0.1), 0 \le x \le 5\}$ with a high pressure ratio as $PR = 10^6$. The computation advances to $t=0.01$ at $\Delta t = 1.0 \times 10^{-5}$ on 200 grids.

(b) Blast wave

The initial condition is $(\rho, u, p) = \{(1, 0, 1000), 0 \le x < 0.1; (1, 0, 0.01), 0.1 \le x \le 0.9; (1, 0, 100), 0.9 < x \le 1\}$ with a solid-wall boundary condition on the two ends, $x = 0\&1$. The computation advances to $t=0.038$ at $\Delta t = 1.0 \times 10^{-5}$ on 600 grids. A result on 15,000 grids by WENO5-JS is used as the "exact" solution.

(c) Shu–Osher problem

The initial condition is $(\rho, u, p) = \{(3.857143, 2.629369, 10.3333), -5 \le x < -4; (1 + 0.2\sin(5x), 0, 1), -4 < x \le 5\}$. The computation advances to $t=1.8$ at $\Delta t = 0.003$ on 240 grids, and the result of WENO5-JS on 10,000 grids is regarded as the "exact" solution.

In the above problems, TVD-RK3 is used for the temporal discretization, and the Steger–Warming scheme is used for flux splitting. Usually, characteristic variables are used to mitigate the oscillation of results.

(3) 2D Euler/ Navier-Stokes equations

The following problems are tested: the 2D Riemann problem, double-Mach reflection, reflected shock-boundary layer interaction in a shock tube, hypersonic half-cylinder flow, and inviscid sharp double cone flow at $M = 9.59$. The temporal scheme, the flux-splitting scheme, and the employment of characteristic variables are the same as that in 1D Euler equations; in a viscous situation, a fourth-order central discretization is employed to compute viscous terms.

(a) 2D Riemann problem

The problem is defined in the domain $[0,1] \times [0,1]$ with the initial conditions as:

$$(\rho, u, v, p) = \begin{cases} (1.5, 0, 0, 1.5), & 0.8 \le x \le 1, 0.8 \le y \le 1 \\ (0.5323, 1.206, 0, 0.3), & 0 \le x < 0.8, 0.8 \le y \le 1 \\ (0.138, 1.206, 1.206, 0.029), & 0.8 \le x < 0.8, 0 \le y < 0.8 \\ (0.5323, 0, 1.206, 0.3), & 0.8 \le x \le 1, 0 \le y < 0.8 \end{cases}$$

The grid number is $960 \times 960$. The computation advances to $t = 0.8$ at $\Delta t = 0.0001$ with a specific heat ratio of $\gamma = 1.4$.

(b) Double Mach reflection

The problem describes a Mach 10 shock impinging on a wall at an incident angle of 60° on the domain $[0,4] \times [0,1]$ with grids of $1920 \times 480$. The initial condition is $(\rho, u, v, p) = \{(8, 7.145, -4.125, 116.5), x < 1/6 + y/\sqrt{3};\ (1.4, 0, 0, 1), x \geq 1/6 + y/\sqrt{3}\}$. The computation runs until $t = 0.2$ at $\Delta t = 0.0001$ with $\gamma = 1.4$.

(c) Reflected shock-boundary layer interaction in a shock tube

As described in [17], the problem is about the evolution of two gases initially separated in the middle of the 2D insulated square tube with unit side length. The initial states of gases are $(\rho, u, v, p) = \{(120, 0, 0, 120/\gamma), 0 < x < 0.5;\ (1.2, 0, 0, 1.2/\gamma), 0.5 \leq x < 1\}$, where $\gamma = 1.4$ and $Pr = 7.3$. In addition, a constant non-dimensional viscosity is chosen as $\mu = 1$ [17]. Due to the symmetry, only half of the tube in the vertical direction is considered, and the corresponding domain is $[0,1] \times [0, 0.5]$. Uniform grids are used in the computation numbering 501×251; especially, the employment of $Re = 200$ is considered in this study. The computation runs to $t = 1$ at $\Delta t = 0.00025$.

(d) Hypersonic half-cylinder flow

Although supersonic half-cylinder flow is supposed to be a trivial test for shock-capturing schemes, cases with hypersonic inflow are seldom tested by improvements of WENO3-Z, and, accordingly, robustness deficiencies may be concealed. In view of this, hypersonic cases with Mach numbers reaching 20 are tested, and a quarter cylinder is chosen due to the symmetry. The grid, numbering 60×60, are shown in Fig. 2.

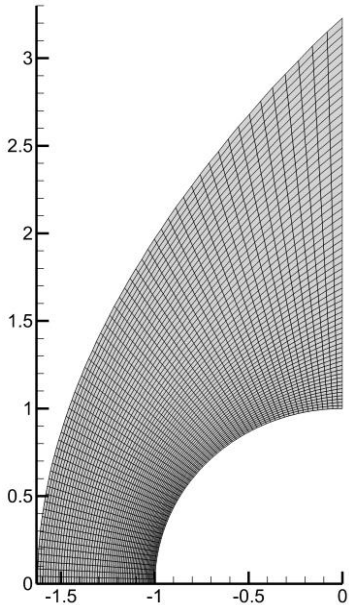
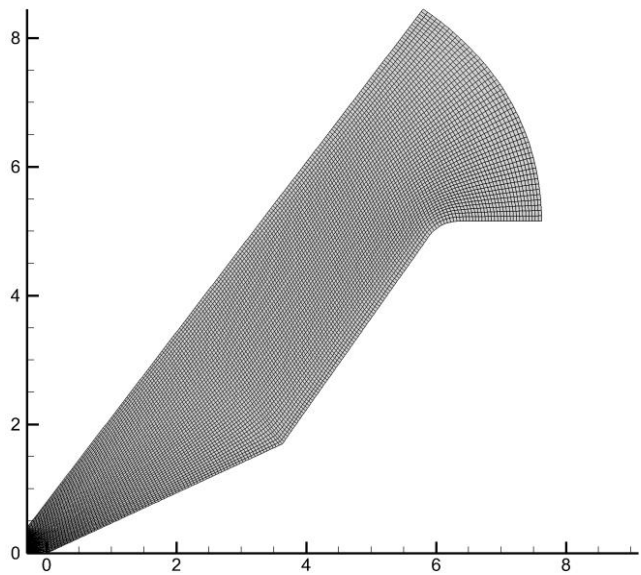

Fig. 2. Grid of the quarter cylinder numbering 60×60

Fig. 3. Grids of the 25°/55° sharp double cone numbering 204×48

(e) Inviscid sharp double cone flow at $Ma = 9.59$

This problem describes a hypersonic, inviscid flow around a 25°/55° double cone. With the apex at the origin, the first deflection is at $x = 3.63$, and the second corner is at $x = 6$. A short extension is configured ahead of the apex with a length of 0.25. The grid, numbering 204×48, are illustrated in Fig. 3.

### 4.2 1D scalar advection equation with the initial condition Eq. (24)

Although the improvements in Table 1 claimed to achieve the third-order at $CP_1$, we determined that [11] the statement did not hold when $CP_1$ occurred at the middle of the grid cell under the $L_\infty$-norm. The performances of the proposed WENO-$Z_{ES2}$ and -$Z_{ES3}$ in such a situation are worthy of study, naturally, and the comparison is made with that of WENO-F3. The WENO-F3 scheme has the smallest $c_{\tau_2}$ among the improvements in Table 1 using $\tau$ in Eq. (9.2), which is thought to be the least dissipative. Using the configuration in "(1)" in Section 4.1, the computation advances until $t = 2$ with $CFL = 0.25$. The corresponding results are shown in Table 2, which manifest the capability of the proposed schemes and the incapability of WENO-F3 in achieving the third-order.

Table 10 $L_\infty$-norm errors and orders of WENO3-$Z_{ES2}$, -$Z_{ES3}$, and WENO-F3 by using the equation of 1D scalar advection with the initial condition Eq. (24) at $t = 2$ and $CFL = 0.25$

| N | Δt | WENO-F3 | | WENO3-$Z_{ES2}$ | | WENO3-$Z_{ES3}$ | |
|---|---|---|---|---|---|---|---|
| | | $L_\infty$-error | $L_\infty$-order | $L_\infty$-error | $L_\infty$-order | $L_\infty$-error | $L_\infty$-order |
| 10 | 0.05 | 2.531E-01 | -- | 2.7305E-01 | -- | 2.9232E-01 | -- |
| 20 | 0.025 | 5.267E-02 | 2.265 | 6.0777E-02 | 2.168 | 7.6272E-02 | 1.938 |
| 40 | 0.0125 | 7.131E-03 | 2.885 | 1.0277E-02 | 2.564 | 1.5602E-02 | 2.289 |
| 80 | 0.00625 | 1.022E-03 | 2.802 | 1.0360E-03 | 3.310 | 8.0086E-03 | 0.962 |
| 160 | 0.003125 | 1.651E-04 | 2.631 | 1.2817E-04 | 3.015 | 1.2820E-04 | 5.965 |
| 320 | 0.0015625 | 3.011E-05 | 2.455 | 1.6035E-05 | 2.999 | 1.6035E-05 | 2.999 |
| 640 | 0.00078125 | 7.368E-06 | 2.031 | 2.0047E-06 | 3.000 | 2.0047E-06 | 3.000 |

For cases where $CP_1$ does not lie on the middle of the grid cell, WENO-$Z_{ES2}$ and -$Z_{ES3}$ achieve the third $L_\infty$-order as expected, which is omitted here for brevity.

### 4.3 1D Euler equations

WENO3-Z and WENO-F3 are chosen for comparison, where the former is regarded as the basic reference and the latter as the representative of improvements to WENO3-Z with high resolution.

(1) Strong shock wave

The density distributions of the proposed schemes are shown in Fig. 4. All schemes fulfill the test, which indicates the capability to solve a strong shock with a high pressure rate. Through the enlarged window of the figure, the proposed schemes indicate relatively better agreement with the exact solution in comparison with WENO3-Z and WENO-F3.

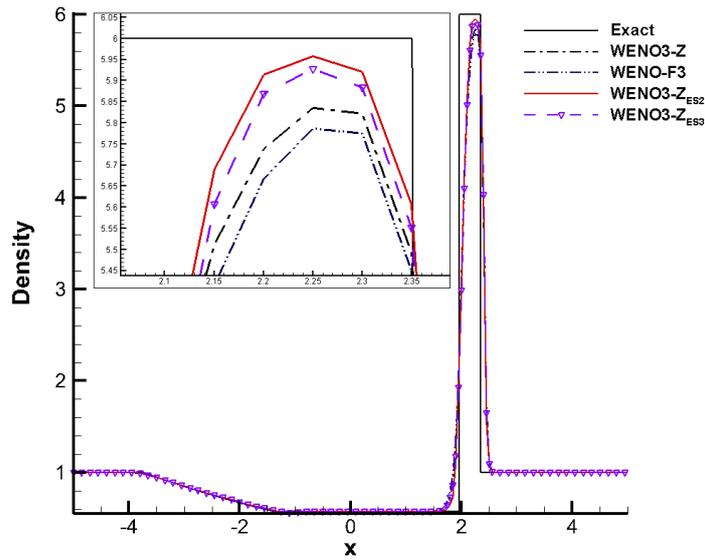

Fig. 4. Density distributions of strong shock waves at $t$=0.01 on 200 grids with an initial pressure ratio of $PR=10^6$ by using WENO-$Z_{ES2}$ and -$Z_{ES3}$, in comparison with WENO3-Z and WENO-F3.

(2) Blast wave

The density distributions of the test schemes are shown in Fig. 5. All schemes accomplish the computation as well. Comparatively, the proposed schemes perform slightly better on resolving the density dip around $x = 0.746$ compared with WENO3-Z and WENO-F3.

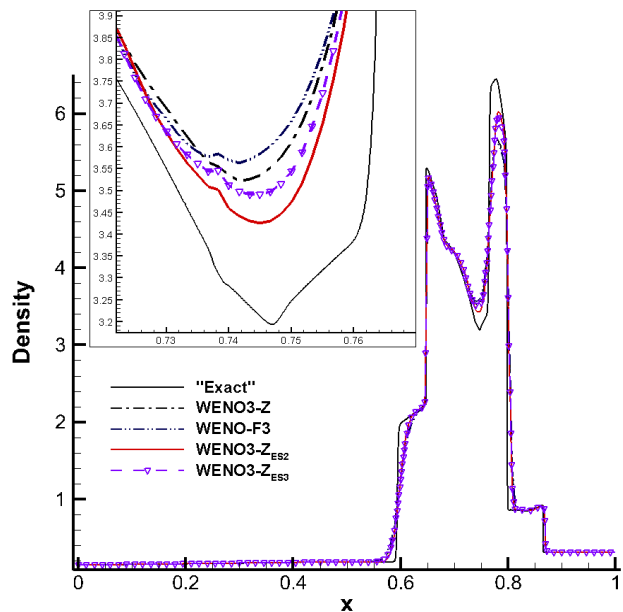

Fig. 5. Density distributions of blast waves at $t$=0.038 on 600 grids by using WENO-$Z_{ES2}$ and -$Z_{ES3}$, in comparison with WENO3-Z and WENO-F3.

(3) Shu–Osher problem

Unlike the typical grid numbers applied by third-order WENO schemes for this problem,

namely 400–600, only 240 grids are used in current study. Unsurprisingly, WENO3-Z barely resolves the 5 peaks/valleys on the density distribution; WENO-F3 shows some improvement but is quite limited. By contrast, the proposed schemes indicate substantially enhanced resolution, where WENO3-$Z_{ES2}$ slightly outperforms WENO3-$Z_{ES3}$.

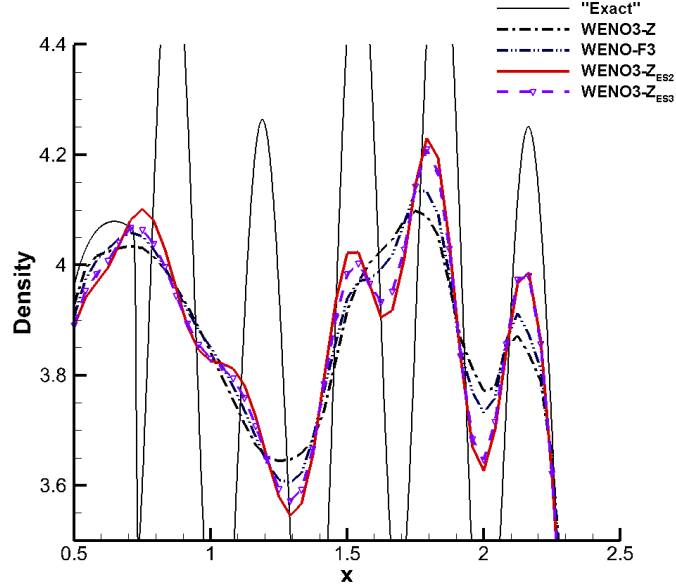

Fig. 6. Density distributions of Shu–Osher problem at $t$=1.8 on 240 grids by using WENO-$Z_{ES2}$ and –$Z_{ES3}$, in comparison with WENO3-Z and WENO-F3.

4.4 2D Euler/Navier-Stokes equations

In 2D problems, WENO3-Z is chosen as the basic reference, while WENO-F3 is the representative improvement for comparison as before.

(1) 2D Riemann problem

The density contours of four schemes are shown in Fig. 7. WENO3-Z yields a "clean" result but with less resolution. For example, the typical instabilities along the slip line are not resolved, whereas the other schemes yield regular roll-ups without obvious numerical noise; further, the latter resolve much more subtleties in the region indicated by the red box in Fig. 7(a). Despite the qualitative similarity of performances among the latter schemes, one obvious distinction exists: WENO3-$Z_{ES2}$ and -$Z_{ES3}$ yield symmetric structures in the aforementioned region with respect to the domain diagonal, which is theoretically plausible and agrees with that of WENO3-Z, whereas WENO-F3 yields asymmetric structures therein. The result indicates the proposed schemes manifest satisfactory numerical stability while maintaining high resolution. As we stated previously, only considering resolution would be inappropriate because the ensuing solutions might lack robustness, which is necessary in engineering. The subsequent "(4)" and "(5)" will address this aspect.

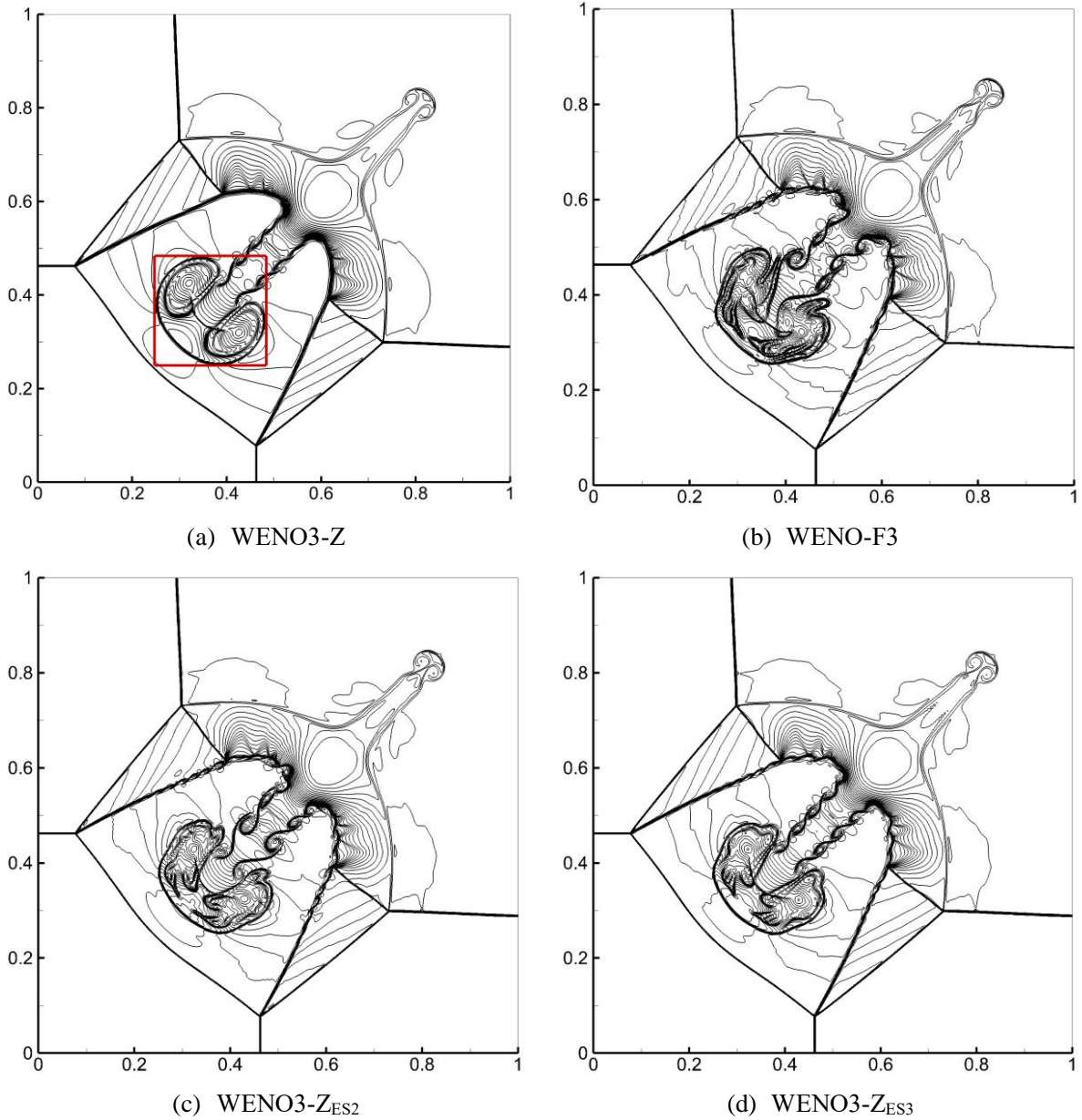

Fig. 7. Density contours of 2D Riemann problem by using WENO3-$Z_{ES2}$ and -$Z_{ES3}$ in comparison with those of WENO3-Z and WENO-F3 on $960 \times 960$ grids at $t = 0.8$ and $\Delta t = 0.0001$ (40 contours from 0.14 to 1.7)

(2) Double Mach reflection

As shown in Fig. 8, all schemes yield smooth distributions except in the region where the slip line occurs. It is a reminder that although many simulations claim to have high resolution on the one hand, a large amount of numerical oscillations can occur in the space after the shocks on the other hand. Regarding the resolution, WENO-$Z_{ES2}$ yields relatively more ripples along the slip line, and WENO-$Z_{ES3}$ ranks second, slightly outperforming WENO-F3.

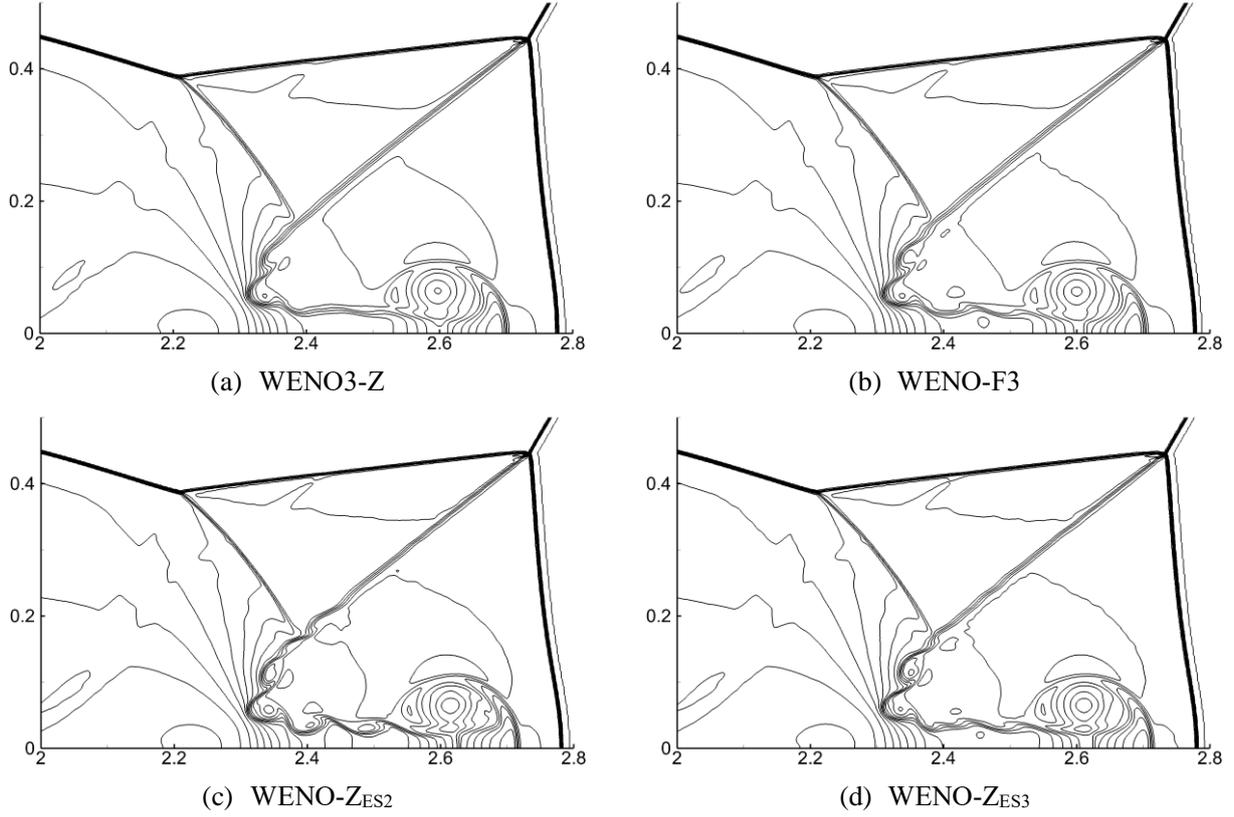

(a) WENO3-Z  (b) WENO-F3

(c) WENO-$Z_{ES2}$  (d) WENO-$Z_{ES3}$

Fig. 8. Density contours of double Mach reflection by using WENO3-$Z_{ES2}$ and -$Z_{ES3}$ in comparison with those of WENO3-Z and WENO-F3 on $1920 \times 480$ grids at $t = 2$ and $\Delta t = 0.0001$ (33 contours from 1.4 to 24)

(3) Reflected shock-boundary layer interaction in a shock tube

In this test, a reflected shock in a shock tube interacts with a boundary and induces three large vortices; schemes with different resolutions will yield different heights of the second vortex. For illustration, the height regarding WENO3-Z can be indicated and measured by the dashed line in Fig. 9(a). From the figure, although four schemes fulfill the computation, the improvements to WENO3-Z yield higher heights of the second large vortex, namely, $\{h_{WENO3-Z}, h_{WENO-F3}, h_{WENO3-Z_{ES2}}, h_{WENO3-Z_{ES3}}\} = \{0.152, 0.166, 0.162, 0.159\}$, where the latter three manifests their superior resolutions. Although WENO-F3 demonstrates a relatively higher height in this case, the scheme indicates less robustness or stability in the following two cases.

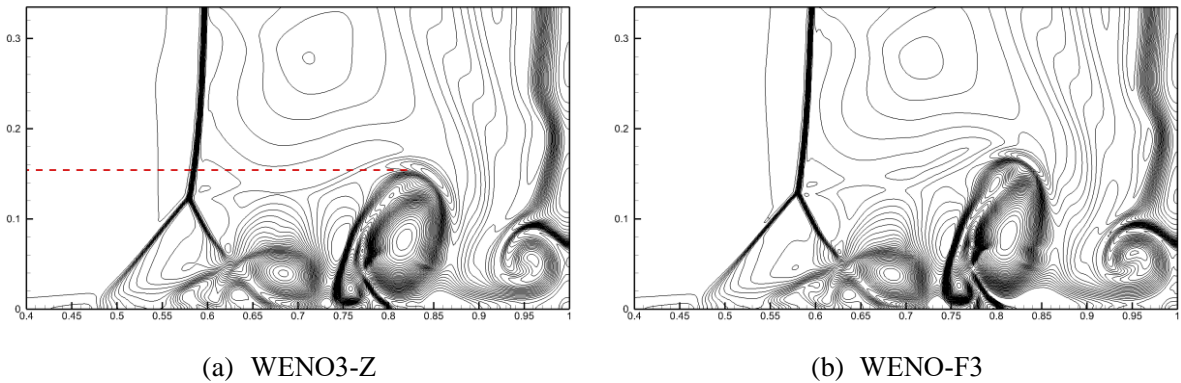

(a) WENO3-Z  (b) WENO-F3

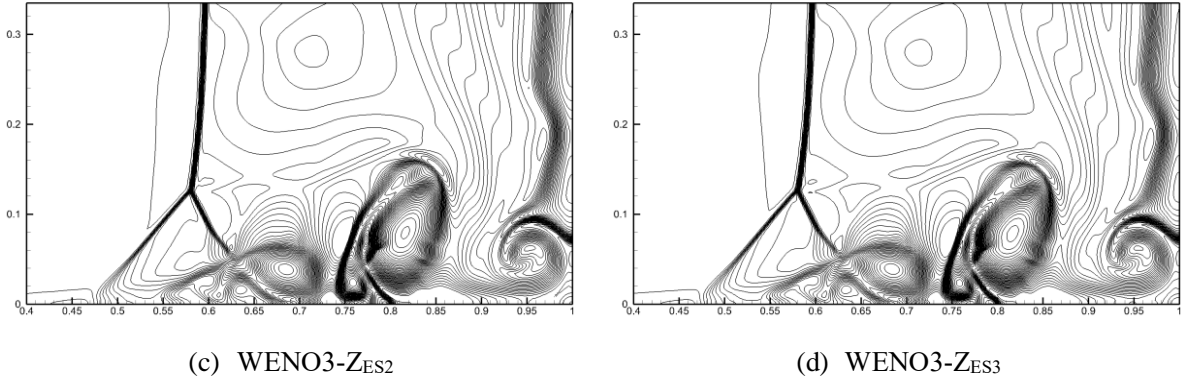

(c) WENO3-$Z_{ES2}$  (d) WENO3-$Z_{ES3}$

Fig. 9. Density contours of reflected shock-boundary layer interaction in a shock tube by using WENO3-$Z_{ES2}$ and -$Z_{ES3}$ in comparison with those of WENO3-Z and WENO-F3 on $501 \times 251$ grids at $t = 1$ and $\Delta t = 0.00025$ (41 contours from 20 to 110)

(4) Hypersonic half-cylinder flow

Although supersonic half-cylinder flow is a small case for shock-capturing schemes, hypersonic tests are seldom reported by improvements aiming for high resolutions. To this end, the computations on hypersonic half-cylinder are carried out where $\Delta t = 0.001$ and 50000 run steps are chosen to ensure a steady solution. Under the current configuration, the approximate upper-limit Mach number at which the test scheme can accomplish the computation is acquired by using one as the increment of Mach number. It is found that the corresponding Mach number limit of WENO3-$Z_{ES2}$ is 16, and that of WENO3-$Z_{ES3}$ is 19, while the isobar contours shown in Fig. 10 indicate rather smooth distributions. By comparison, the limit of WENO-F3 is obtained as 12, which is smaller than the above two. For reference, the limit of WENO3-Z in this case is found to be $M = 19$, and the corresponding result is shown in Fig. 10(a), with high-quality smoothness and shock-capturing manifested. In short, considering previous studies on resolutions, WENO3-$Z_{ES2}$ and WENO3-$Z_{ES3}$ represent sufficient robustness while possessing high resolution.

As previously mentioned, if $\tau_4$ in WENO3-$Z_{ES3}$ is replaced by $\left|\delta_j^{(2)_2}\delta_j^{(3)_1}\right|$, the upper-limit Mach number will decrease to 14, which numerically exhibits the robustness of the former indicator.

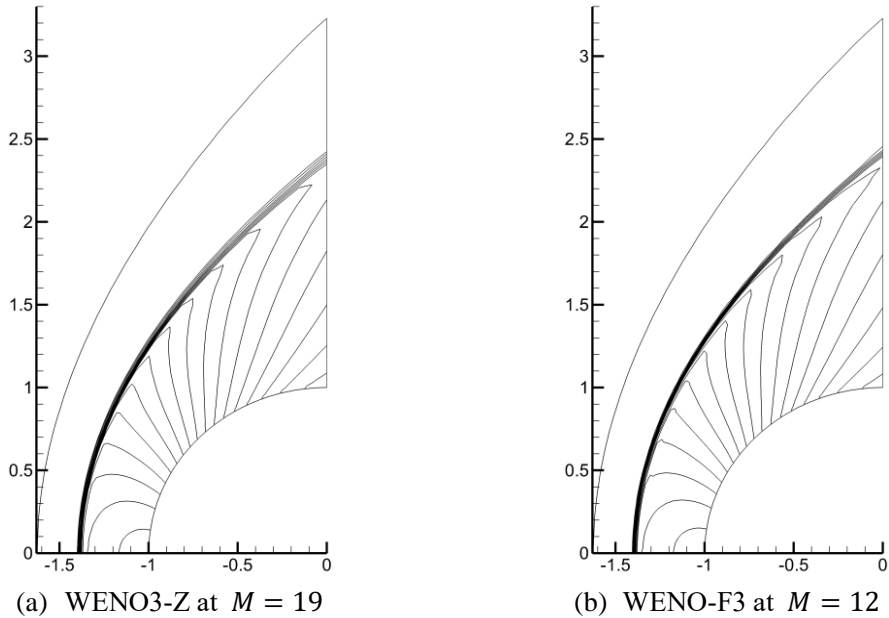

(a) WENO3-Z at $M = 19$  (b) WENO-F3 at $M = 12$

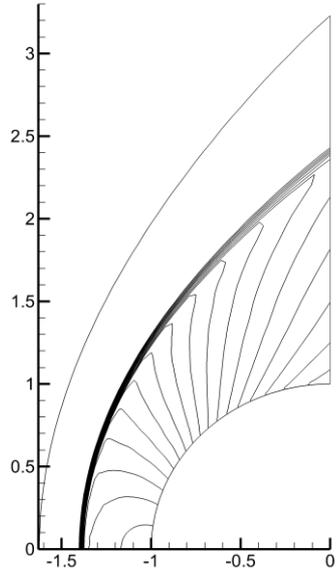
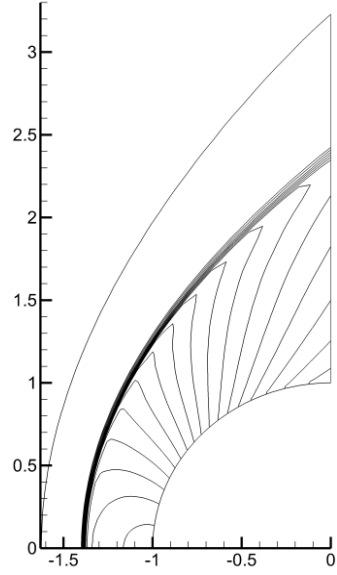

(c) WENO3-$Z_{ES2}$ at $M = 16$　　　　(d) WENO3-$Z_{ES3}$ at $M = 19$

Fig. 10. Pressure contours of hypersonic half-cylinder flows by using WENO3-$Z_{ES2}$ and -$Z_{ES3}$ in comparison with those of WENO3-Z and supersonic results of WENO-F3 on 60×60 grids at $t = 50$ and $\Delta t = 0.001$ (20 contours from 0.05 to 21)

(5) Inviscid sharp double cone flow at $M = 9.59$

　　The pressure contours of WENO3-Z are given in Fig. 11(a) as a reference, where tiny oscillations are indicated ahead of the reflected shocks. Also shown in Figs. 11(c)-(d), WENO3-$Z_{ES2}$ and -$Z_{ES3}$ manifest their capability of shock-capturing with smooth variable distributions, which even outperform that of WENO3-Z somewhat. Although WENO-F3 fulfills the computation, the scheme yields obvious oscillations ahead of the shock waves, as shown in Fig. 11(b); moreover, quantitative investigation shows that the oscillations therein yield an overshot Mach number of 12.26. Hence, the superior robustness of the proposed schemes is indicated.

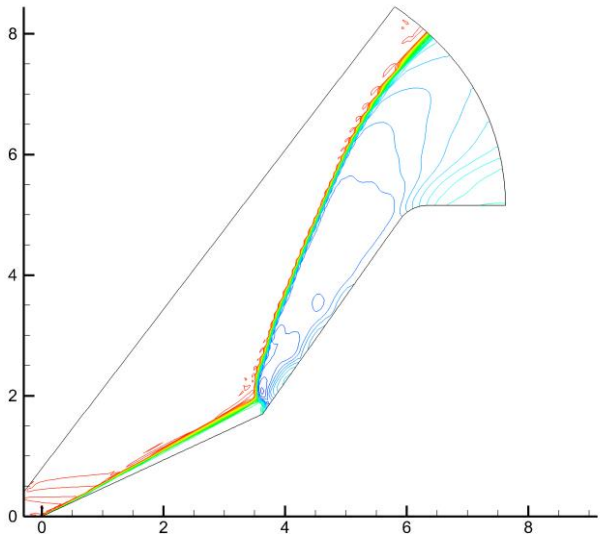
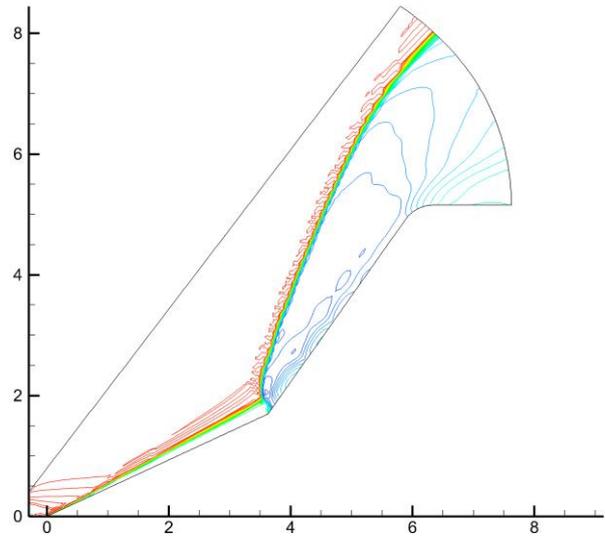

(a) WENO3-Z　　　　(b) WENO-F3

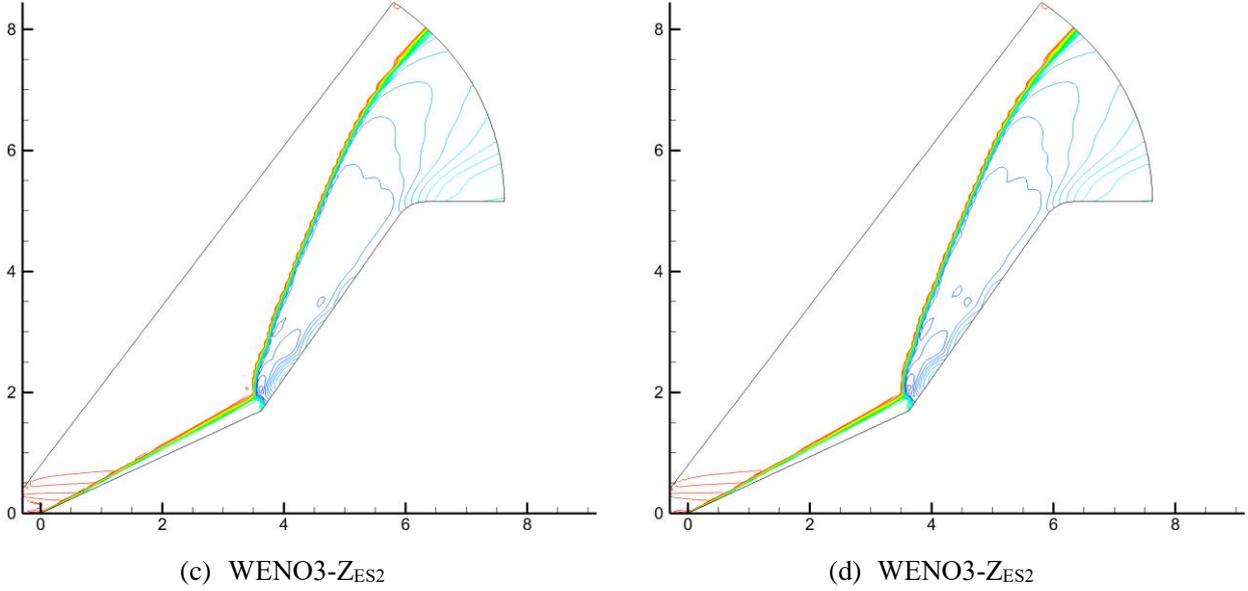

(c) WENO3-Z$_{ES2}$          (d) WENO3-Z$_{ES2}$

Fig. 11. Mach number contours of inviscid sharp double cone flow at $M = 9.59$ by using WENO3-Z$_{ES2}$ and -Z$_{ES3}$ in comparison with those of WENO3-Z and WENO-F3 on 120×60 grids at $t = 250$ and $\Delta t = 0.005$ (38 contours from 0 to 9.591)

## 5 Conclusions and discussions

Under the context of third-order scale-independent WENO-Z type schemes, we investigate improvements to WENO3-Z to enhance robustness while achieving optimal order at $CP_1$. The following conclusions are drawn:

(1) Although there are many improvements aiming to achieve the third-order at $CP_1$, such as those in Table 1, they barely fulfill the job when $CP_1$ occurs at the middle of grid cells due to not considering the occurrence of critical points within grid intervals. We once [11] proposed a solution by devising a $\tau$ with an error order of 7 in such a situation combined with the extension of the original $\beta_1^{(2)}$, but the corresponding scheme indicated inferior robustness at large Mach numbers. A similar lack of robustness exists for improvements to WENO3-Z.

(2) By heuristic analysis and numerical tests, we suppose the too-large error order of $\tau$ is unfavorable for numerical stability, based on which a new idea is proposed: on the one hand, a new $\beta_k^{(2)*}$ is devised that is defined on an expanded stencil having the error order of 2 or 4 in the absence or presence of $CP_1$ and performs more stably; on the other hand, a new $\tau_4$ is devised that has an error order of 4 or 5 in the absence or presence of $CP_1$. Rigorous analysis guarantees $\alpha_k$ and subsequently $\omega_k$ satisfy the sufficient condition to achieve the third-order in a scale-independent manner. Consequently, two improvements, WENO3-Z$_{ES2}$ and -Z$_{ES3}$, are developed, and their achievement of optimal order at $CP_1$ is theoretically indicated and numerically validated.

(3) Numerical tests show that WENO3-Z$_{ES2}$ and -Z$_{ES3}$ indicate an enhanced robustness (e.g., the accomplishment of computations on hypersonic half-cylinder flow with Mach numbers equal to or larger than 16, and the smooth solutions of inviscid sharp double cone flow at $M = 9.59$), which contrasts the comparative improvement to WENO3-Z. Meanwhile, the proposed schemes show high resolutions also, especially in 1D Shu–Osher problems.

Future studies will focus on the improvement of efficiency as described previously, as well as the further development of symmetric schemes in favor of higher resolution.

## Acknowledgment

This study was sponsored by a project of the National Numerical Wind Tunnel of China under grant number NNW2019ZT4-B12.